\documentclass[preprint,10pt]{aastex}

\singlespace

\providecommand{\eg}{e.g.}
\providecommand{\etal}{et~al.}
\providecommand{\ie}{i.e.}

\received{}
\revised{}
\accepted{}

\slugcomment{To appear in \apj}

\shorttitle{GRB 081029}
\shortauthors{S.~T. Holland}

\begin{document}


\title{GRB 081029: A Gamma-Ray Burst with a Multi-Component Afterglow}

\author{Stephen~T.~Holland\altaffilmark{1,2,3},
        Massimiliano~De~Pasquale\altaffilmark{4},
        Jirong~Mao\altaffilmark{5,6,7,8},
        Takanori~Sakamoto\altaffilmark{1,9,3},
        Patricia~Schady\altaffilmark{10,4},
        Stefano~Covino\altaffilmark{5},
        Yi-Zhong~Fan\altaffilmark{11},
        Zhi-Ping~Jin\altaffilmark{5,11},
        Paolo~D'Avanzo\altaffilmark{5},
        Angelo~Antonelli\altaffilmark{12},
        Valerio~D'Elia\altaffilmark{12},
        Guido~Chincarini\altaffilmark{5},
        Fabrizio~Fiore\altaffilmark{12},
        Shashi~Bhushan~Pandey\altaffilmark{13,14}, \&
        Bethany~E.~Cobb\altaffilmark{15}
       }

\altaffiltext{1}{Astrophysics Science Division, Code 660.1,
                 8800 Greenbelt Road
                 Goddard Space Flight Centre,
                 Greenbelt, MD 20771
                 U.S.A.
                 \email{Stephen.T.Holland@nasa.gov}}

\altaffiltext{2}{Universities Space Research Association
                 10211 Wincopin Circle, Suite 500
                 Columbia, MD 21044
                 U.S.A.}

\altaffiltext{3}{Centre for Research and Exploration in Space Science and Technology
                 Code 668.8
                 8800 Greenbelt Road
                 Goddard Space Flight Centre,
                 Greenbelt, MD 20771
                 U.S.A.}

\altaffiltext{4}{Mullard Space Science Laboratory,
                 University College London,
                 Holmbury St Mary,
                 Dorking Surrey RH5~6NT,
                 UK}

\altaffiltext{5}{INAF-Osservatorio Astronomico di Brera,
                 Via Emilio Bianchi 46,
                 I--23807 Merate (LC),
                 Italy}

\altaffiltext{6}{Yunnan Observatory,
                 Chinese Academy of Sciences,
                 Kumming, Yunan Province, 650011,
                 China}

\altaffiltext{7}{International Centre for Astrophysics,
                 Korea Astronomy and Space Science Institute 776,
                 Daedeokdae-ro, Yuseong-gu, Daejeon,
                 Republic of Korea 305--348}

\altaffiltext{8}{Key Laboratory for the Structure and Evolution of Celestial Objects,
                 Chinese Academy of Sciences, Kunming, Yunnan Province, 650011,
                 China}

\altaffiltext{9}{Joint Centre for Astrophysics,
                 University of Maryland, Baltimore County,
                 1000 Hilltop Circle,
                 Baltimore, MD 21250,
                 USA}

\altaffiltext{10}{Max-Planck Institut f{\"u}r Extraterrestrische Physik,
                  Giessenbachstra{\ss}e,
                  D--85748 Garching,
                  Germany}

\altaffiltext{11}{Purple Mountain Observatory,
                  Chinese Academy of Sciences,
                  Nanjing 210008,
                  China}

\altaffiltext{12}{INAF-Osservatorio Astronomico di Roma,
                  Via de Frascati 33,
                  I--00040 Monteporzio Catone (Roma),
                  Italy}

\altaffiltext{13}{Randall Laboratory of Physics,
                  University of Michigan,
                  450 Church St,
                  Ann Arbor, MI 48109--1040,
                  USA} 

\altaffiltext{14}{Aryabhatta Research Institute of Observational Sciences,
                  Manora Peak,
                  Nainital, 263129,
                  India}

\altaffiltext{15}{Department of Physics,
                  The George Washington University,
                  725 21st St NW,
                  Washington, DC 20052,
                  USA}


\begin{abstract}

  We present an analysis of the unusual optical light curve of the
  gamma-ray burst GRB~081029, a long--soft burst with a redshift of $z
  = 3.8479$.  We combine X-ray and optical observations from the {\sl
    Swift\/} X-Ray Telescope and the {\sl Swift\/} UltraViolet/Optical
  Telescope with ground-based optical and infrared data obtained using
  the REM, ROTSE, and CTIO 1.3-m telescopes to construct a detailed
  data set extending from 86~s to $\sim$100\,000~s after the BAT
  trigger.  Our data cover a wide energy range, from 10~keV to 0.77~eV
  (1.24~{\AA} to 16\,000~{\AA}).  The X-ray afterglow shows a shallow
  initial decay followed by a rapid decay starting at about 18\,000~s.
  The optical and infrared afterglow, however, shows an
  uncharacteristic rise at about 3000~s that does not correspond to
  any feature in the X-ray light curve.  Our data are not consistent
  with synchrotron radiation from a jet interacting with an external
  medium, a two-component jet, or continuous energy injection from the
  central engine. We find that the optical light curves can be broadly
  explained by a collision between two ejecta shells within a
  two-component jet.  A growing number of gamma-ray burst afterglows
  are consistent with complex jets, which suggests that some (or all)
  gamma-ray burst jets are complex and will require detailed modelling
  to fully understand them.

\end{abstract}




\section{Introduction\label{SECTION:intro}}

There is a great deal of variety in the observed optical and infrared
light curves of gamma-ray burst (GRB) afterglows.  Most exhibit some
form of power-law decay as predicted by a model with synchrotron
emission from the forward shock of ejecta ploughing into an external
medium \citep{R1999,SPH1999}.  However afterglow light curves often
display rises, flares, breaks, and other behaviour that require
extensions to this simple picture.  See \citep[\eg,][]{PV2011} for a
detailed discussion of the variety seen in GRB optical afterglows.
\citet{OPS2009} find that before about 500~s light curves can either
rise or decay from the first observation, but after about 500~s a
significant fraction of afterglows decay with a power law.  In general
the only significant temporal evolution after this time is one or more
breaks in the power-law decay.  There are, however, several GRBs that
have had optical afterglows that exhibit significant rebrightening or
flaring after about 500~s.  Some examples are GRB~970508, which
brightened by about one magnitude after about one day \citep{DMK1997},
GRB~060614, which peaked at about six hours \citep{DCP2006,MHM2007},
and GRB~100418A, which peaked at about 14 hours \citep{MAB2011}.  One
problem with studying unusual afterglows is that the observation
density is often not great enough to resolve rapid changes in the
optical properties of afterglows.

GRB~081029 was detected by the BAT at 01:43:56 UT on 2008 Oct 29.  The
{\sl Swift\/} observatory \citep{GCC2004} is a multi-instrument
satellite mission that was designed to detect and rapidly localize
GRBs.  The observatory contains three telescopes.  The Burst Alert
Telescope \citep[BAT;][]{BBC2005} is used to identify GRBs and
localize them to $\sim$3$\arcmin$ in the energy range 15--150 keV.
Once BAT has localized a burst {\sl Swift\/} slews to point the X-Ray
Telescope \citep[XRT;][]{BHN2005} and the UltraViolet/Optical
Telescope \citep[UVOT;][]{RKM2005} at the burst.  The XRT obtains
X-ray localizations to $\lesssim$5$\arcsec$ in the energy range
0.2--10 keV while the UVOT obtains localizations to $\sim 0\farcs5$,
then cycles through a set of optical and ultraviolet filters covering
the wavelength range from 1700~{\AA} to 6500~{\AA}.  {\sl Swift\/} was
unable to slew immediately to this burst due to an Earth limb
constraint, so the first {\sl Swift\/} narrow-field observations did
not begin until approximately 45 minutes after the BAT trigger.  The
BAT light curve showed a single smooth peak and had a $T_{90}$
duration of $270 \pm 45$~s \citep{CBB2008}.  The spectrum of the
prompt emission was well-fit by a simple power law.

The ROTSE-IIIc telescope located the optical afterglow of GRB~081029
86~s after the burst \citep{R2008}.
The afterglow was also detected in the infrared by the REM telescope
at 154~s \citep{CCA2008}, but not in the ultraviolet with {\sl
  Swift\/}/UVOT \citep{HS2008}.  UVOT optical data showed a rise
between approximately 2700 and 9000 s while \citet{C2008} detected the
afterglow in both the optical and infrared using ANDICAM on the CTIO
1.3-m telescope.  They found that the afterglow decayed with a
power-law index of approximately $0.9$ between about 9000 and
13\,000~s after the trigger.  Further early observations were reported
by PROMPT starting 92~s after the trigger \citep{WHB2008} as well as
by GROND 8 minutes after the trigger \citep{CLG2008,NGK2011}.  The XRT
found a fading source \citep{GOB2008} at the ROTSE-IIIc location
\citep{R2008}.
The Australia Compact Telescope Array observed GRB~081029
approximately one month after the burst at 4.800 and 4.928 GHz, but
did not detect the afterglow.  Their merged data at 4.800 and
4.928~GHz yielded a radio flux density, at the afterglow position, of
$f_\nu = -0.168 \pm 0.219$~mJy per beam \citep{MTP2008}.

A redshift of $z = 3.8479 \pm 0.0002$ was measured from several
absorption features by the VLT/UVES \citep{DCD2008} and was confirmed
by Gemini-South/GMOS \citep{CFC2008}.  The VLT/UVES spectrum is
presented in \S~\ref{SECTION:spectrum}.  The GMOS spectrum shows
evidence for a damped Lyman-alpha system as well as several metal
absorption features in the host galaxy.

GRB~081029 was unusual even amongst the GRBs with unusual optical light
curves.  The initial light curve decayed in the normal way, but there
was a sudden increase in flux at about 3000~s \citep{NGK2011} that
cannot be explained using the convention afterglow model.  In this
paper we present space- and ground-based gamma-ray, X-ray,
ultraviolet, optical, and near-infrared observations of GRB~081029.
We will propose that the X-ray, optical, and infrared data suggest
that the afterglow of GRB~081029 can broadly be explained by the
collision of a fast-moving ejecta shell with a slower shell within a
two-component jet.

We present our data in \S~\ref{SECTION:data} and give the results of
our analyses of the spectral energy distribution (SED) and light
curves in \S~\ref{SECTION:results}.  We explore various scenarios to
explain the rapid brightening of GRB~081029's afterglow in
\S~\ref{SECTION:interp}.


\section{Data\label{SECTION:data}}

\subsection{BAT Data\label{SECTION:bat_data}}

\def\eiso{E_{\rm iso}}
\def\egamma{E_{\gamma}}
\def\ep{E_{\rm peak}}
\def\epo{E^{\rm obs}_{\rm peak}}
\def\eps{E^{\rm src}_{\rm peak}}
 
The BAT data analysis was performed using the {\sl Swift\/} HEASOFT
6.5.1 software package.  The burst pipeline script, {\sc
  batgrbproduct}, was used to process the BAT event data.  We used the
the position of the optical afterglow as the source's input position
during the process.

\begin{figure}
  \centerline{
  \includegraphics[scale=0.7,angle=-90]{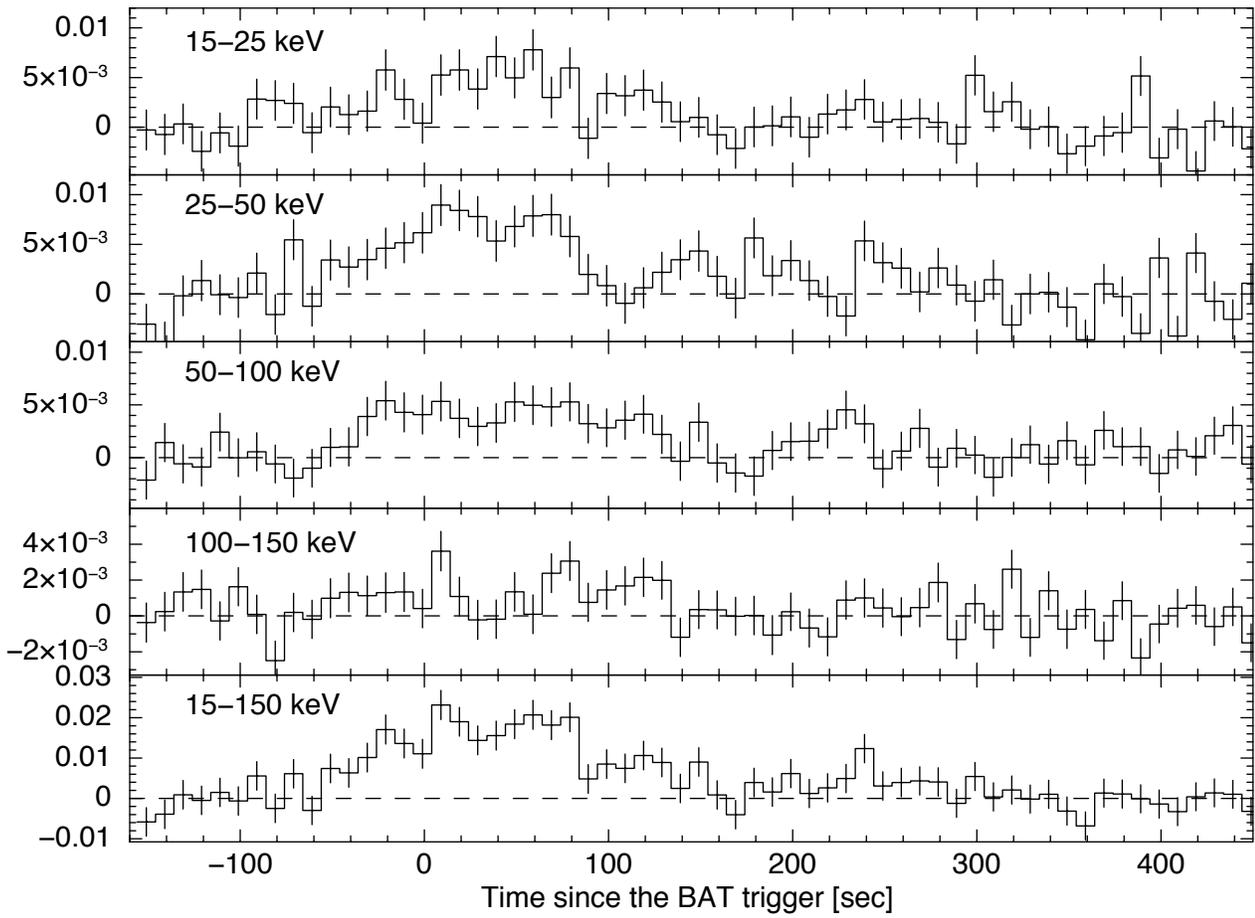}}
  \caption{The BAT energy-resolved light curves of GRB~081029 with
    10~s binning.\label{FIGURE:bat_lc}}
\end{figure}

Figure~\ref{FIGURE:bat_lc} shows the BAT energy-resolved light curves
of GRB~081029 with 10~s binning.  The light curve shows an extremely
weak and smooth profile with a $T_{90}$ duration of $280 \pm 50$~s
(1~$\sigma$, statistical).  The 1~s peak flux in the 15--150 keV band
measured in the 1~s time window starting from 20.6~s after the BAT
trigger time is $(2.8 \pm 1.3) \times 10^{-8}$ erg~cm$^{-2}$~s$^{-1}$.
The energy fluence in the 15--150 keV band is $(2.0 \pm 0.2) \times
10^{-6}$ erg~cm$^{-2}$.  The time-average spectrum is well fitted by a
simple power law with the photon index of $1.5 \pm 0.2$.

\begin{figure}
  \centerline{
    \includegraphics[scale=0.7,angle=-90]{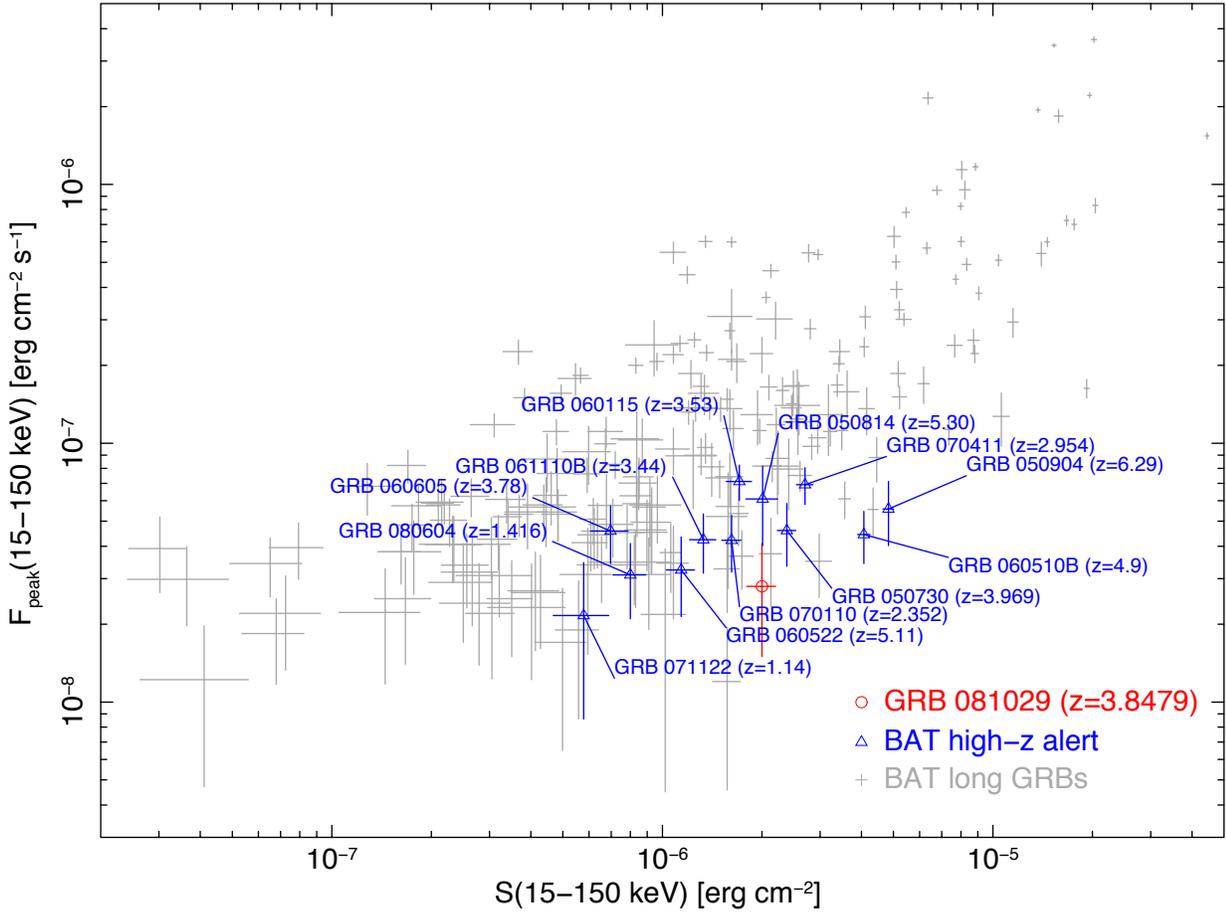}}
  \figcaption{Distribution of the 1~s peak energy flux in the 15--150
    keV band versus the energy fluence in the 15--150 keV band.
    GRB~081029 is indicated with a red circle.  BAT GRBs with known
    redshifts that satisfy the \citet{USS2008} high-$z$ criteria are
    indicated with blue triangles, and BAT long GRBs are shown with
    gray crosses.\label{FIGURE:bat_ene_fluence_flux}}
\end{figure}

Because of the weak, smooth light curve of the prompt emission,
GRB~081029 satisfies the BAT possible high-$z$ criteria
\citep{USS2008}.  The BAT possible high-$z$ criteria are basically
selecting those bursts with weak, smooth light curves and hard
spectra.
Figure~\ref{FIGURE:bat_ene_fluence_flux} shows the distributions of
GRB~081029, the BAT known-$z$ bursts which satisfy the \citet{USS2008}
BAT possible high-$z$ criteria, and the BAT long GRBs in the peak flux
and the fluence plane.  The BAT parameters are from the BAT1 catalogue
\citep{SBB2008a}.  As seen in the figure, GRB~081029 has a lower peak
flux and fluence than the higher redshift bursts such as GRB~050904,
GRB~060510B, and GRB~050814.  We also note that some very faint GRBs,
such as GRB~071122 and GRB~080604 occurred at low redshifts.
Therefore, the weakness and the smoothness of the GRB~081029 hard
X-ray light curve in the prompt emission might be more related to the
central engine of the burst rather than the cosmological redshift
effect.  The \citet{USS2008} test gives a reasonable indication that a
burst may be at high redshifts but the false negatives---such as
GRB~080913A, GRB~090423, GRB~090429B, and GRB~090429B
\citep{GKF2009,TFL2009,ZZV2009,CLF2011}, which were at high redshift
but did not satisfy the criteria---mean that the test should be used
with extreme caution.

\subsection{XRT Data\label{SECTION:xrt_data}}
 
XRT began to observe GRB~081029 2448~s after the BAT trigger.  The
UVOT-enhanced X-ray position is RA, Dec.\ = 23:07:05.51, $-$68:09:21.9
(J2000.0) with an uncertainty of $1\farcs5$ (radius, 90\% confidence).
The observational data were processed by the {\sl Swift\/} Data Center
at NASA/GSFC and further calibrated with {\sc xrtpipeline}.
For details of how the light curve was produced see \citet{EBP2007}.
All the XRT data for GRB~081029 were collected in Photon Counting
mode.

\begin{figure}
  \includegraphics[scale=0.6,angle=-90]{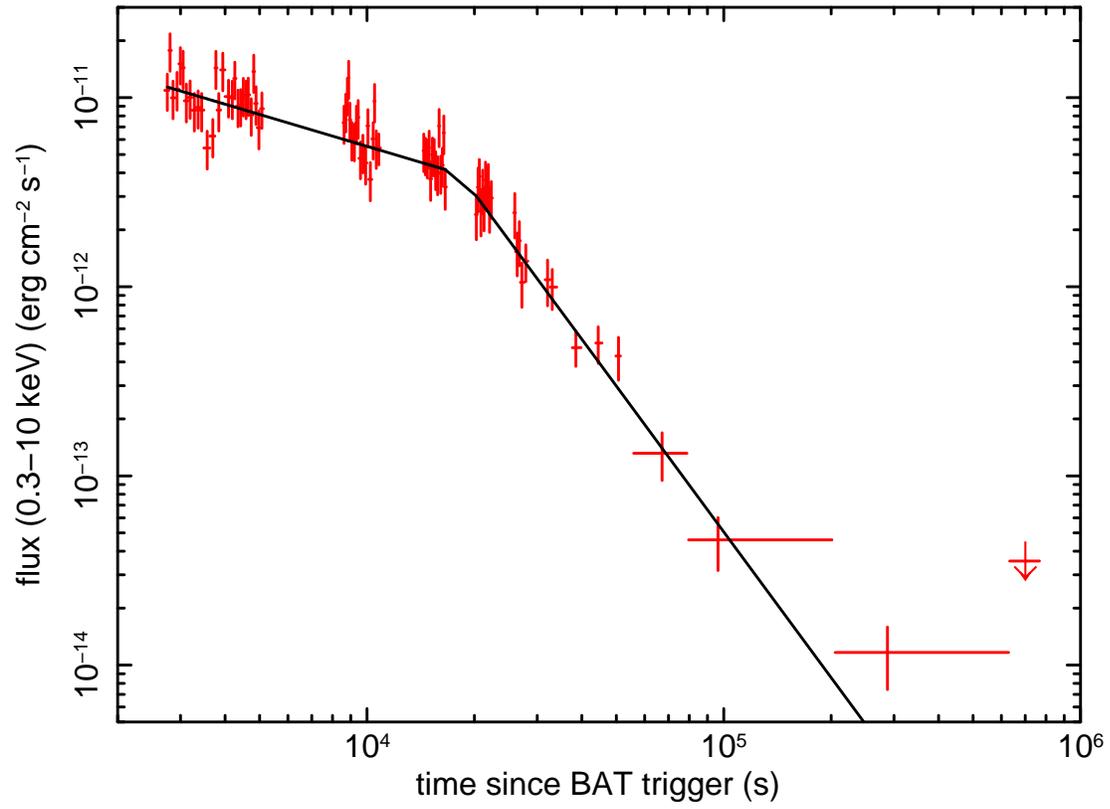}
  \figcaption[./xrt.eps]{The Figure shows the {\sl Swift\/}/XRT light
    curve. The data have not been corrected for Galactic absorption or
    absorption in the host galaxy.\label{FIGURE:xrtlc}}
\end{figure}

The X-ray light curve can be modelled by a broken power law ($f(t)
\propto t^{-\alpha}$).  The best-fitting model has decay indices of
$\alpha_{X,1} = 0.56 \pm 0.03$ and $\alpha_{X,2} = 2.56 \pm 0.09$ with
a break time of $t_{X,b} = 18\,230 \pm 346$~s yielding a
goodness-of-fit of $\chi^2/\mathrm{dof} = 93.947/77 = 1.22$.  The
X-ray light curve with this fit is shown in
Figure~\ref{FIGURE:xrtlc}.  Alternately, if we fit a smoothly-varying
broken power law \citep{BHR1999} we find $\alpha_{X,1} = 0.45 \pm
0.11$, $\alpha_{X,2} = 2.65 \pm 0.23$, and a smoothness parameter of
$n = 2.3 \pm 1.5$ with $\chi^2/\mathrm{dof} = 91.260/76 = 1.20$.  The
initial X-ray light curve shows some evidence for flaring between
approximately 2500 and 5000~s after the BAT trigger.  The sawtooth
behaviour of the X-ray emission during this period is consistent
with flares with $\Delta t/t \la 1$.  It is possible that the X-ray
photons that we see at this time are due to flaring on top of a power
law decay.  The lack of X-ray data before 2000~s could be causing us
to miss the rise of the flare and thus give the impression that the
X-ray photons seen between 2000~s and 5000~s are due solely to the
plateau phase of the X-ray light curve.

The X-ray spectrum can be fit by an absorbed power law with a photon
index $\Gamma_X = 1.98 \pm 0.08$, corresponding to $\beta_X = 0.98 \pm
0.08$.  The assumed Galactic column density value in the direction of
the burst is $N_H = 2.8 \times 10^{20}$~cm$^{-2}$ \citep{KBH2005} and
the fitted intrinsic column density in the host galaxy is $N_H =
4.9^{+3.9}_{-2.7} \times 10^{21}$~cm$^{-2}$.  Assuming an SMC-like
relation between the neutral hydrogen column density and the $V$-band
extinction of $N_H = (15.4 \times 10^{21}) A_V$ \citep[equation~(4)
and Table~2 of][]{P1992} this corresponds to $A_V =
0.3^{+0.3}_{-0.2}$~mag in the rest frame of the host galaxy.  However,
the observed gas-to-dust ratio for GRB host galaxies varies by about a
factor of ten \citep{SMP2007}, so the X-ray data alone can only
constrain the rest frame $V$-band extinction to be $A_V \lesssim
2$~mag.
The observed 0.3--10 keV flux is $3.1 \times
10^{-12}$~erg~cm$^{-2}$~s$^{-1}$, which corresponds to an unabsorbed
value of $\sim 3.5 \times 10^{-12}$~erg~cm$^{-2}$~s$^{-1}$.  This was
computed using the time-average spectrum between $2.7 \times 10^3$~s
and $6.2 \times 10^4$~s after the BAT trigger.


\subsection{UVOT Data\label{SECTION:uvot_data}}

\begin{figure}
  \includegraphics[scale=0.6,angle=+90]{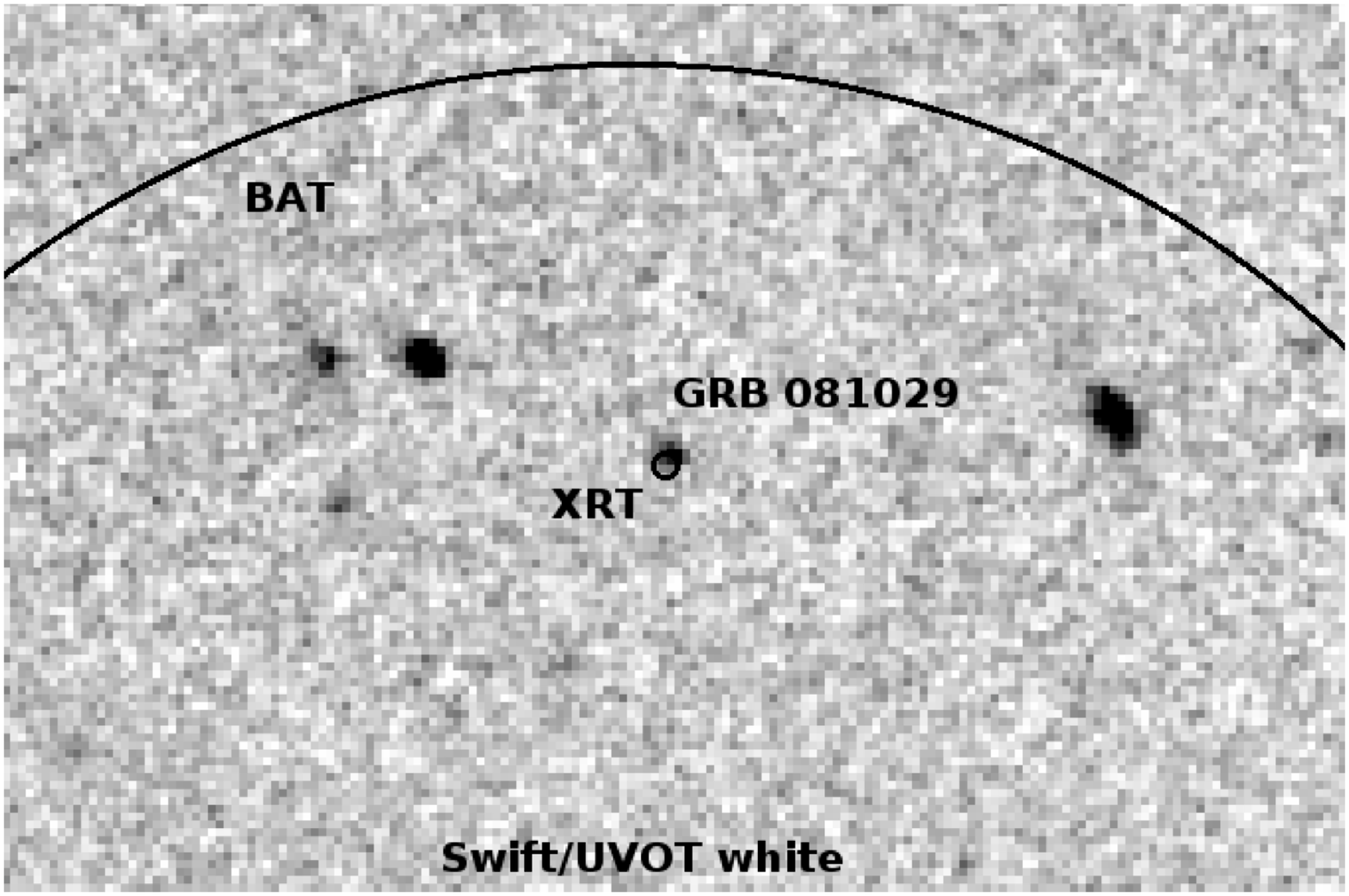}
  \caption{The Figure shows a {\sl Swift\/}/UVOT {\sl white\/} image
    of the field of GRB~081029.  The BAT and XRT error circles are
    shown.  The XRT error circle has a radius of $1\farcs5$.  North is
    up and east is to the left.\label{FIGURE:uvot}}
\end{figure}

The {\sl Swift\/}/UVOT began settled observations 2708 s after the BAT
trigger \citep{SBB2008b}.  An optical afterglow was detected in the
initial white exposure with a magnitude of $20.47^{-0.29}_{+0.39}$.
The afterglow increased in luminosity until approximately 9000 s and
then faded.  The UVOT position of the afterglow is RA, Dec.\ =
23:07:05.34, $-$68:09:20.0 with an estimated internal uncertainty of
$0\farcs14$ and an estimated systematic uncertainty relative to the
ICRS \citep{FMA2004} of $0\farcs42$ \citep{BCH2010}.  These
uncertainties are the 90\% confidence intervals.  This corresponds to
Galactic coordinates of $\ell^\mathrm{II},b^\mathrm{II} =
316\fdg5827,-46\fdg1091$.
The field of GRB~081029 is shown in Figure~\ref{FIGURE:uvot}.  The
afterglow is well-isolated from other sources in the field, so there
is no contamination from neighbouring sources when doing aperture
photometry.

\begin{figure}
  \plotone{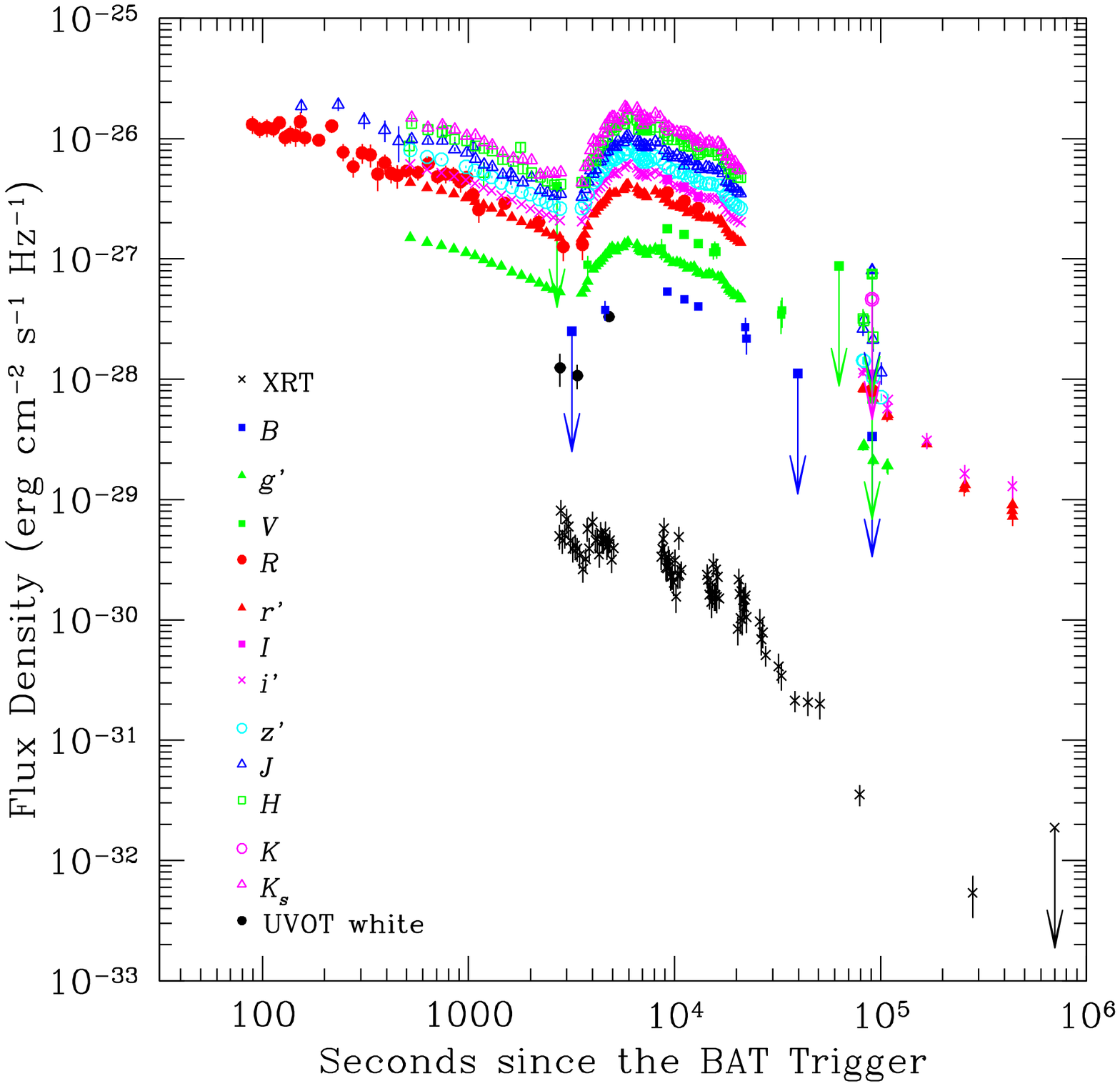} \figcaption[./flux_density.eps]{This
    Figure shows the flux-density light curves for the X-ray data
    (black crosses) and the optical and infrared data.  The
    optical/infrared photometry has not been corrected for Galactic
    extinction along the line of sight to the burst, and the $X$-ray
    data have not been corrected for Galactic absorption.  The
    vertical bars represent one-sided, 1-$\sigma$ error bars in the
    flux density.  In most cases the error bars are smaller than the
    plotting symbol.\label{FIGURE:flux}}
\end{figure}

We obtained the UVOT data from the {\sl Swift\/} Data
Archive\footnote{The {\sl Swift\/} Data Archive is hosted by
  HEASARC.}.  These data have had bad pixels identified, mod-8 noise
corrected, and have been transformed into FK5 coordinates.  We used
the standard UVOT data analysis software distributed with HEASOFT 6.10
along with version 20110131 of the UVOT calibration data.  Photometry
was done using {\sc uvotsource} with circular aperture of radius
$2\farcs5$ and a nearby circular background region with a radius of
$10\arcsec$.
The background region was selected to have similar background
properties to those at the location of the afterglow, and to be free
of contaminating sources.  The UVOT photometry is presented in
Table~\ref{TABLE:photometry}.  The photometry was calibrated to the
UVOT photometric system described in \citet{PBP2008,BLH2011}.  We have
followed the \citet{PBP2008} convention and used lowercase letters to
identify the UVOT bandpasses.  Figure~\ref{FIGURE:flux} shows the UVOT
light curves for filters where a detection was found.

\begin{deluxetable}{rcrcc}
\tabletypesize{\scriptsize}
\tablewidth{0pt}
\tablecaption{The {\sl Swift\/}/UVOT photometry of GRB~081029.  These
  data have not been corrected for either Galactic extinction or any
  possible extinction in the host galaxy.  The first column is the
  midpoint time of the observation in seconds since the BAT trigger
  (2008 Oct 29 at 01:43:56 UT).  The second column is the filter name
  while the third column is the total exposure time in seconds.  The
  fourth column gives the magnitude, or the 3-$\sigma$ upper limit if
  there was no detection.  The fifth column gives the 1-sided
  1-$\sigma$ statistical errors in the
  magnitude.\label{TABLE:photometry}}

\tablehead{%
        \colhead{Time (s)} &
        \colhead{Filter} &
        \colhead{Exposure (s)} &
        \colhead{Magnitude} &
        \colhead{Error}}
\startdata
    2783 & white &     147 &     20.47 & $-0.19, +0.39$ \\
    3377 & white &     197 &     20.64 & $-0.22, +0.28$ \\
    4812 & white &     197 &     19.41 & $-0.10, +0.10$ \\
101\,891 & white &    3219 &  $>$23.2  & \\

    3787 & $v$   &     197 &     19.02 & $-0.18, +0.22$ \\
    8657 & $v$   &     197 &     18.69 & $-0.20, +0.25$ \\
 15\,407 & $v$   &     295 &     18.76 & $-0.13, +0.15$ \\
 15\,710 & $v$   &     295 &     18.66 & $-0.12, +0.14$ \\
 16\,014 & $v$   &     295 &     18.75 & $-0.13, +0.15$ \\
 32\,933 & $v$   &     295 &     20.06 & $-0.29, +0.39$ \\
 33\,237 & $v$   &     295 &     19.98 & $-0.27, +0.37$ \\
 62\,874 & $v$   &      44 &  $>$19.1  & \\
102\,333 & $v$   &    2996 &  $>$21.8  & \\
222\,873 & $v$   &    9630 &  $>$22.2  & \\
312\,805 & $v$   &    9418 &  $>$22.2  & \\
498\,648 & $v$   &    6804 &  $>$21.8  & \\
596\,971 & $v$   & 11\,369 &  $>$22.1  & \\
639\,792 & $v$   & 17\,900 &  $>$22.4  & \\
726\,351 & $v$   & 14\,756 &  $>$22.2  & \\

    3172 & $b$   &     197 &  $>$20.5  & \\
    4607 & $b$   &     197 &     20.08 & $-0.18, +0.22$ \\
 22\,088 & $b$   &     295 &     20.44 & $-0.19, +0.23$ \\
 22\,352 & $b$   &     219 &     20.67 & $-0.25, +0.32$ \\
 39\,624 & $b$   &     295 &  $>$21.4  & \\
101\,443 & $b$   &    3217 &  $>$22.2  & \\
 
    2967 & $u$   &     197 &  $>$20.0  & \\
101\,188 & $u$   &     321 &  $>$20.3  & \\

    4197 & uvw1  &     197 &  $>$20.4  & \\

    3992 & uvm2  &     197 &  $>$20.0  & \\

    3582 & uvw2  &     197 &  $>$20.4  & \\
\enddata
\end{deluxetable}

\subsection{Ground-Based Data\label{SECTION:ground_data}}

\subsubsection{REM Data\label{SECTION:rem_data}}

Observations of the afterglow of GRB~081029 were carried out with the
REM telescope \citep{ZCG2001,CZA2003,CSS2004} equipped with the ROSS
optical spectrograph/imager and the REMIR near-infrared camera on 2008
Oct 29, starting about 154 seconds after the burst \citep{CCA2008}.
The night was clear, with a seeing of about $2\farcs0$.  We collected
images with typical exposure times from 5 to 120 seconds, covering a
time interval of about 0.5 hours.  The complete observing log is
presented in Table~\ref{TABLE:tab_log_REM}.

Image reduction was carried out by following the standard procedures:
subtraction of an averaged dark frame, then division by a normalized
flat.  For the near-infrared data an average sky value was subtracted
before dividing by the flat field.  Astrometry was performed using the
USNO-B1.0\footnote{\tt http://www.nofs.navy.mil/data/fchpix/} and the
2MASS\footnote{\tt http://www.ipac.caltech.edu/2mass/} catalogues.  We
performed aperture photometry with the {\sc SExtractor} package
\citep{BA1996} for all the objects in the field.  In order to minimize
any systematic effect, we performed differential photometry with
respect to a selection of local isolated and non-saturated comparison
stars.  The $J$ and $H$ data were reduced following the method
described in \citet{DAP2002}.  The near-infrared photometry was
calibrated against the 2MASS catalogue.  Given the non-photometric
conditions during the observing night optical imaging was
cross-calibrated against the SMARTS photometry, which was obtained
under better sky conditions (see \S~\ref{SECTION:andicam}) by
selecting a common set of bright, non-saturated field stars.

\begin{deluxetable}{cccccc}
\tabletypesize{\scriptsize}
\tablewidth{0pt}
\tablecaption{REM observation log for GRB~081029.  Magnitudes are not
  corrected for reddening.\label{TABLE:tab_log_REM}}
\tablehead{%
  \colhead{Mid Obs.\ Time (UT)} &
  \colhead{$t-t_0$ (s)} &
  \colhead{Exposure (s)} &
  \colhead{Instrument} &
  \colhead{Mag} &
  \colhead{Filter}}
\startdata
20081029.07814 &    515    &    23.0                  & REM/REMIR   & $15.18 \pm 0.30$   & $H$ \\
20081029.08122 &    781    &    35.5                  & REM/REMIR   & $14.86 \pm 0.16$   & $H$ \\
20081029.08590 &   1185    &    61.0                  & REM/REMIR   & $15.73 \pm 0.24$   & $H$ \\
20081029.09291 &   1791    &    86.0                  & REM/REMIR   & $15.21 \pm 0.12$   & $H$ \\

20081029.07396 &    154    &    36.0                  & REM/REMIR   & $14.83 \pm 0.13$   & $J$ \\
20081029.07488 &    233    &    35.5                  & REM/REMIR   & $14.80 \pm 0.12$   & $J$ \\
20081029.07579 &    312    &    35.5                  & REM/REMIR   & $15.12 \pm 0.17$   & $J$ \\
20081029.07671 &    391    &    35.5                  & REM/REMIR   & $15.33 \pm 0.22$   & $J$ \\
20081029.07748 &    458    &    23.5                  & REM/REMIR   & $15.56 \pm 0.36$   & $J$ \\
20081029.08437 &   1053    &    61.0                  & REM/REMIR   & $15.67 \pm 0.17$   & $J$ \\
20081029.09079 &   1608    &    85.5                  & REM/REMIR   & $16.23 \pm 0.21$   & $J$ \\

20081029.07600 &  330.048  & $1 \times 30.0 $	      & REM/ROSS    & $16.42 \pm 0.16$   & $R$ \\
20081029.07646 &  369.792  & $1 \times 30.0 $	      & REM/ROSS    & $16.75 \pm 0.21$   & $R$ \\
20081029.07692 &  409.536  & $1 \times 30.0 $	      & REM/ROSS    & $16.89 \pm 0.24$   & $R$ \\
20081029.07737 &  448.416  & $1 \times 30.0 $	      & REM/ROSS    & $16.83 \pm 0.22$   & $R$ \\
20081029.07782 &  487.296  & $1 \times 30.0 $	      & REM/ROSS    & $16.78 \pm 0.22$   & $R$ \\
20081029.07828 &  527.040  & $1 \times 30.0 $	      & REM/ROSS    & $16.91 \pm 0.24$   & $R$ \\
20081029.07873 &  565.920  & $1 \times 30.0 $	      & REM/ROSS    & $16.95 \pm 0.25$   & $R$ \\
20081029.07918 &  604.800  & $1 \times 30.0 $	      & REM/ROSS    & $17.11 \pm 0.29$   & $R$ \\
20081029.07964 &  644.544  & $1 \times 30.0 $	      & REM/ROSS    & $16.98 \pm 0.26$   & $R$ \\
20081029.08009 &  683.424  & $1 \times 30.0 $	      & REM/ROSS    & $16.88 \pm 0.24$   & $R$ \\
20081029.08054 &  722.304  & $1 \times 30.0 $	      & REM/ROSS    & $16.87 \pm 0.23$   & $R$ \\
20081029.08101 &  762.912  & $1 \times 30.0 $	      & REM/ROSS    & $16.94 \pm 0.25$   & $R$ \\
20081029.08146 &  801.792  & $1 \times 30.0 $	      & REM/ROSS    & $16.89 \pm 0.24$   & $R$ \\
20081029.08191 &  840.672  & $1 \times 30.0 $	      & REM/ROSS    & $17.11 \pm 0.29$   & $R$ \\
20081029.08253 &  894.240  & $1 \times 60.0 $	      & REM/ROSS    & $17.16 \pm 0.16$   & $R$ \\
20081029.08334 &  964.224  & $1 \times 60.0 $	      & REM/ROSS    & $17.14 \pm 0.16$   & $R$ \\
20081029.08414 & 1033.344  & $1 \times 60.0 $	      & REM/ROSS    & $17.48 \pm 0.22$   & $R$ \\
20081029.08494 & 1102.464  & $1 \times 60.0 $	      & REM/ROSS    & $17.44 \pm 0.21$   & $R$ \\
20081029.08609 & 1201.824  & $1 \times 120.0$	      & REM/ROSS    & $17.38 \pm 0.12$   & $R$ \\
20081029.08758 & 1330.560  & $1 \times 120.0$	      & REM/ROSS    & $17.49 \pm 0.13$   & $R$ \\
20081029.09108 & 1632.960  & $1 \times 60.0 $	      & REM/ROSS    & $17.50 \pm 0.22$   & $R$ \\
20081029.09187 & 1701.216  & $1 \times 60.0 $	      & REM/ROSS    & $17.55 \pm 0.23$   & $R$ \\
20081029.09269 & 1772.064  & $1 \times 60.0 $	      & REM/ROSS    & $17.69 \pm 0.26$   & $R$ \\
20081029.09383 & 1870.560  & $1 \times 120.0$	      & REM/ROSS    & $17.88 \pm 0.18$   & $R$ \\
20081029.09532 & 1999.296  & $1 \times 120.0$	      & REM/ROSS    & $17.99 \pm 0.20$   & $R$ \\
\enddata
\end{deluxetable}

We detect the optical and near-infrared afterglow identified by
\citet{R2008} and \citet{CLG2008} in our first $H$- and $R$-band
images at the following coordinates: RA, Dec.\ = 23:07:05.33,
$-$68:09:20.0 (J2000.0) with a 1-$\sigma$ error of $0\farcs3$.
Our data are given in Table~\ref{TABLE:tab_log_REM} and shown in
Figure~\ref{FIGURE:flux}.  The power-law decay indices are $\alpha_H =
0.21 \pm 0.41$, $\alpha_J = 0.54 \pm 0.07$, and $\alpha_R = 0.63 \pm
0.07$ in the $H$, $J$, and $R$-band, respectively.

\subsubsection{ROTSE Data\label{SECTION:rotse_data}}

GRB~081029 was observed by ROTSE-IIIc located at the H.E.S.S\@. site
at Mt.\ Gamsberg, Namibia, several times over approximately 17~hr
starting 86.0~s after the BAT trigger \citep{R2008}.  The ROTSE
observations were taken unfiltered, but the CCD's quantum efficiency
peaks at about the $R$ band and the magnitudes were calibrated against
USNO-B stars. Therefore, the ROTSE magnitudes are essentially
equivalent to $R_c$-band magnitudes.  The raw images were processed
using the standard ROTSE software pipeline and photometry was
performed on co-added images using the method described in
\citet{QRY2006}.  Table~\ref{TABLE:tab_log_ROTSE} lists the ROTSE
observations of the afterglow of GRB 081029, and they are plotted in
Figure~\ref{FIGURE:flux}.

\begin{deluxetable}{cccc}
\tabletypesize{\scriptsize}
\tablewidth{0pt}
\tablecaption{ROTSE observation log for GRB~081029.  Magnitudes are not
  corrected for reddening.\label{TABLE:tab_log_ROTSE}}
\tablehead{%
  \colhead{$t-t_0$ (s)} &
  \colhead{Mag} &
  \colhead{$-$Err} &
  \colhead{$+$Err}}
\startdata
   89.220435 & 15.92 & -0.17 & +0.20 \\
   97.120160 & 16.03 & -0.15 & +0.18 \\
  105.019280 & 15.99 & -0.17 & +0.20 \\
  112.919869 & 16.02 & -0.14 & +0.17 \\
  120.819853 & 15.89 & -0.12 & +0.14 \\
  128.719663 & 16.20 & -0.15 & +0.18 \\
  136.619302 & 16.13 & -0.17 & +0.20 \\
  144.619942 & 16.16 & -0.19 & +0.23 \\
  152.619718 & 15.87 & -0.20 & +0.24 \\
  160.518838 & 16.20 & -0.15 & +0.18 \\
  187.520268 & 16.26 & -0.11 & +0.13 \\
  216.818422 & 15.96 & -0.13 & +0.15 \\
  246.219132 & 16.50 & -0.16 & +0.19 \\
  275.619497 & 16.80 & -0.19 & +0.23 \\
  304.420246 & 16.52 & -0.19 & +0.22 \\
  333.519507 & 16.56 & -0.21 & +0.26 \\
  362.620064 & 16.96 & -0.26 & +0.34 \\
  391.818339 & 16.72 & -0.17 & +0.20 \\
  421.219913 & 16.95 & -0.17 & +0.20 \\
  450.419052 & 16.99 & -0.21 & +0.26 \\
  499.120182 & 16.89 & -0.13 & +0.15 \\
  568.520118 & 16.92 & -0.14 & +0.17 \\
  637.619728 & 16.74 & -0.13 & +0.15 \\
  706.619372 & 17.01 & -0.14 & +0.16 \\
  775.719846 & 16.96 & -0.17 & +0.20 \\
  844.720355 & 16.97 & -0.15 & +0.18 \\
  913.719308 & 17.12 & -0.18 & +0.22 \\
  982.318489 & 17.03 & -0.14 & +0.16 \\
 1051.019967 & 17.39 & -0.16 & +0.18 \\
 1120.119318 & 17.70 & -0.22 & +0.27 \\
 1500.119642 & 17.57 & -0.08 & +0.09 \\
 2190.069070 & 17.96 & -0.16 & +0.19 \\
 2880.419956 & 18.47 & -0.23 & +0.29 \\
 3570.918956 & 18.42 & -0.24 & +0.32 \\
60256.035925 & $>$19.61 & \nodata & \nodata \\
\enddata
\end{deluxetable}

\subsubsection{ANDICAM Data\label{SECTION:andicam}}

We obtained four epochs of optical/infrared imaging of the afterglow
of GRB~081029 using the ANDICAM (A Novel Dual Imaging
CAMera)\footnote{\tt http://www.astronomy.ohio-state.edu/ANDICAM}
instrument mounted on the 1.3-m telescope at CTIO\@.  This telescope
is operated as part of the Small and Moderate Aperture Research
Telescope System (SMARTS) consortium\footnote{\tt
  http://www.astro.yale.edu/smarts}.  During each epoch, multiple
short observations were obtained in each filter (45~s in $BI$, 30~s in
$VR$, 15~s in $K$, and 10~s in $JH$) with dithering between
observations via slight telescope offsets and by an internal tilting
mirror system in the infrared.  Standard IRAF data reduction was
performed on these images, including cosmic ray rejection in the
optical images\footnote{via L.A. Cosmic \tt
  http://www.astro.yale.edu/dokkum/lacosmic/} and sky subtraction in
the infrared.  The individual images from each filter were then
aligned and averaged to produce a single frame per epoch.  The SMARTS
observations were conducted such that each of these frames has the
same time of mid-exposure regardless of filter.  The relative
magnitude of the afterglow in each filter was determined by comparison
with a number of nonvariable sources in the GRB~081029 field.  The
relative magnitudes were then converted to true apparent magnitudes
based on the brightness of the afterglow in the first epoch.  Since
all observations were conducted under photometric conditions, the
optical magnitudes of the afterglow in the first epoch were determined
by comparison with Landolt standard stars in the field of T~Phe
\citep{L1992}.  The infrared photometric calibration of the first
epoch was performed using Two Micron All Sky Survey \citep{SCS2006}
stars in the field of GRB~081029.  The afterglow was only
significantly detected in the $R$ and $I$ frames during the
$4^\mathrm{th}$ epoch.  The 3-$\sigma$ limiting magnitudes of the
other images are reported in Table~\ref{TABLE:log_andicam} and plotted
in Figure~\ref{FIGURE:flux}.

\begin{deluxetable}{rcrcrcrcrcrcrcr}
\tabletypesize{\scriptsize}
\tablewidth{0pt}
\tablecaption{ANDICAM observation log for GRB~081029.  Magnitudes are not
  corrected for reddening.\label{TABLE:log_andicam}}
\tablehead{%
  \colhead{$t-t_0$ (s)} &
  \colhead{$B$} &
  \colhead{Err} &
  \colhead{$V$} &
  \colhead{Err} &
  \colhead{$R$} &
  \colhead{Err} &
  \colhead{$I$} &
  \colhead{Err} &
  \colhead{$J$} &
  \colhead{Err} &
  \colhead{$H$} &
  \colhead{Err} &
  \colhead{$K$} &
  \colhead{Err}}

\startdata
 9231 & 19.69 & 0.04 & 18.26 & 0.03 & 17.32 & 0.02 & 16.79 & 0.03 & 15.83 & 0.08 & 15.00 & 0.08 & 14.25 & 0.08 \\
11146 & 19.85 & 0.04 & 18.38 & 0.03 & 17.65 & 0.02 & 17.15 & 0.03 & 16.04 & 0.08 & 15.12 & 0.08 & 14.34 & 0.08 \\
13045 & 20.00 & 0.04 & 18.57 & 0.03 & 17.65 & 0.02 & 17.15 & 0.03 & 16.04 & 0.08 & 15.32 & 0.08 & 14.59 & 0.08 \\ 
90951 & $>$22.7 & \nodata & $>$21.8 & \nodata & 21.43 & 0.02 & 20.84 & 0.03 &
   $>$18.2 & \nodata & $>$17.8 & \nodata & $>$17.8 & \nodata \\
\enddata
\end{deluxetable}

\subsubsection{UVES Spectrum\label{SECTION:spectrum}}

The GRB~081029 optical afterglow was observed with the high resolution
UV-visual echelle spectrograph \citep[UVES;][]{DDK2000}, mounted on
the VLT-UT2 telescope, in the framework of the ESO program 082.A-0755.
Observations began on 2008 Oct 29 at 02:06:37 UT ($\sim 23$~min after
the {\sl Swift\/}/BAT trigger), when the magnitude of the afterglow
was $R \sim 18$.  Two UVES exposures of 5 and 10 minutes were obtained
using both the blue and the red arms.  The slit width was set to
$1\arcsec$ (corresponding to a resolution of $R = 40\,000$) and the
read-out mode was rebinned to $2 \times 2$~pixels.  The spectral range
of our observation is $\sim 3300$~{\AA} to $\sim 9500$~{\AA}.

\begin{figure}
  \plotone{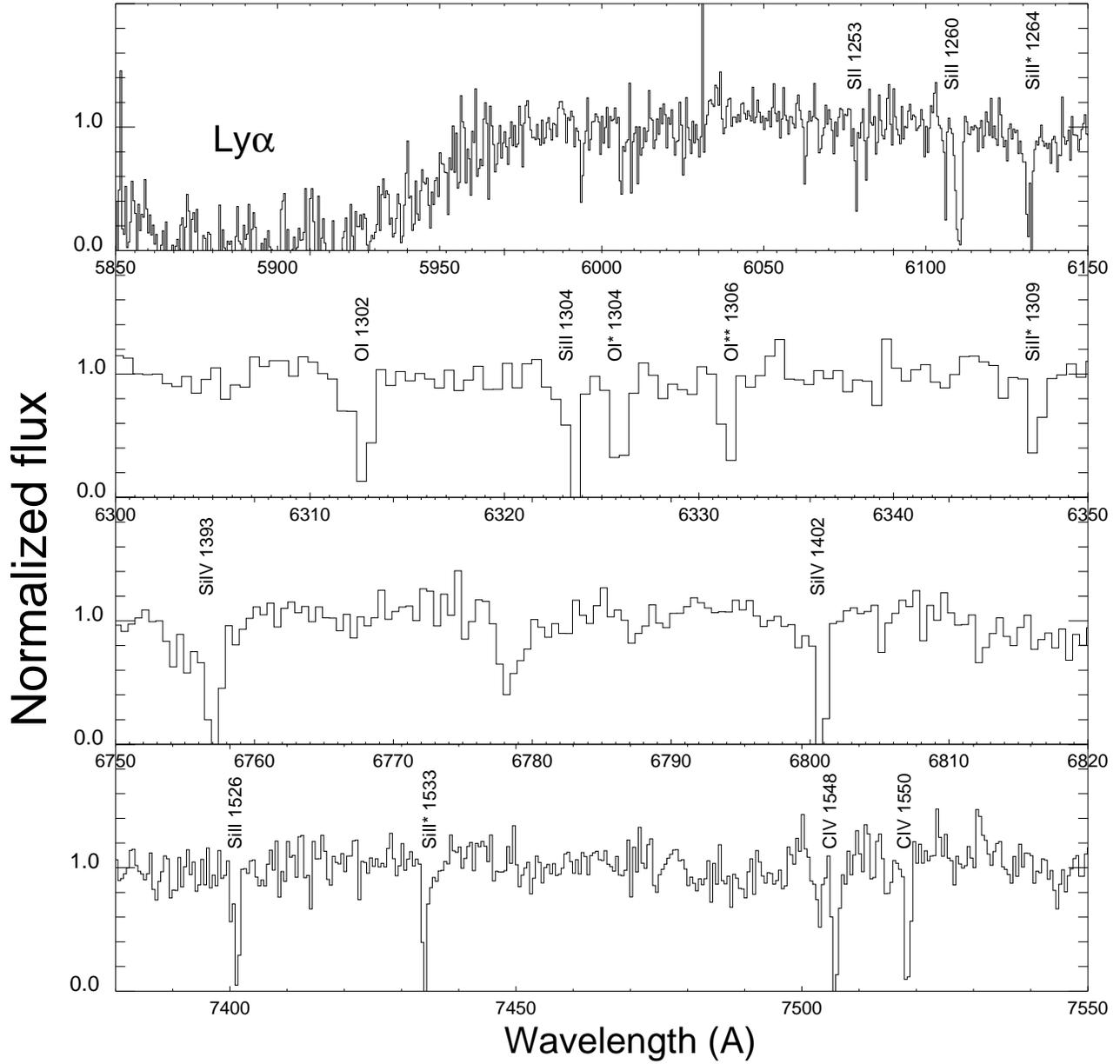} \figcaption[./UVES081029a.ps]{UVES
    spectrum of the optical afterglow of GRB~081029 showing details of
    the absorption system.\label{FIGURE:uves_spectrum}}
\end{figure}

The data reduction was performed using the UVES pipeline
\citep[version 2.9.7;][]{BMB2000}.  Due to the faintness of the
target, the decaying magnitude during the observations, and the
exposure times, the signal-to-noise ratio was not high enough to study
line variability.  The signal-to-noise ratio of the combined spectrum
is $\sim 3$--4, allowing the identification of the main spectral
features, but not a reliable estimation of their column densities.  We
are able to put a 2-$\sigma$ upper limit on the equivalent width of
the intervening \ion{MG}{2}($\lambda$2796) absorption line of
0.6~\AA\@.  A portion of the UVES spectrum with a collection of
absorption features is shown in Figure~\ref{FIGURE:uves_spectrum}.
The redshift path analysed is $1.2 \le z \le 2.7$.

\subsection{Extinction\label{SECTION:extinction}}

The line-of-sight Galactic extinction in the direction of GRB~081029
is $E_{B\!-\!V} = 0.03 \pm 0.01$ mag \citep{SFD1998}.  Using the
extinction law given in \citet{RKO2009} yields extinctions in the UVOT
filters of $A_v = 0.09$, $A_b = 0.12$, $A_u = 0.15$, $A_{uvw1} =
0.20$, $A_{uvm2} = 0.28$, and $A_{uvw2} = 0.25$, and $A_{white} =
0.13$~mag.  For the ground-based data we adopted the \citet{SFD1998}
extinction values of $A_R = 0.08$, $A_{I_C} = 0.06$, $A_J = 0.03$, $A_H
= 0.02$, and $A_K = 0.01$~mag in this direction.


\section{Results\label{SECTION:results}}

\subsection{Spectral Energy Distribution\label{SECTION:sed}}

SEDs for GRB~081029 were produced at three epochs.  The first SED was
constructed for $T+4000$~s using data between 3000~s and 5000~s after
the BAT trigger.  This epoch had X-ray data, GROND data, and UVOT $b$-
and $v$-band data and corresponds to the period when the optical flux
was rising.  The second SED was constructed for $T+12\,000$~s using
data between 9000~s and 14\,000~s.  This epoch had X-ray and the
ground-based data data and corresponds to the period when the optical
and near-infrared fluxes were near their peak.  The final SED was
computed for $T+20\,000$~s using data between 15\,000~s and 25\,000~s
after the BAT trigger.  This epoch had X-ray and ground-based data and
corresponds to the decay after the peak optical/near-infrared flux.
Data were interpolated to a common time within each epoch using the
observed light curves for each filter during the appropriate epoch.

We used {\sc uvot2pha} v1.3 to convert UVOT image data to spectral
files compatible with the spectral fitting package {\sc XSpec}.
Version 104 of the UVOT response matrix calibration was adopted for
the responsivity curves.  For the ground based data, spectral files
were produced for each filter using the appropriate responsivity
curves and setting the magnitude to those determined from the light
curve interpolations. $R$ and $I$ responsivity curves were taken from
\citet{B1990}, and the $J$, $H$ and $K$ band responsivity curves were
taken from \citet{CWB1992,CWW1992} and \citet{BCP1998}.  The GROND
filter response functions\footnote{\tt
  http://www.mpe.mpg.de/\~{}jcg/GROND/GROND\_filtercurves.txt} were
used for the GROND data.

XRT spectra were extracted within {\sc xselect} (v2.4) over the
0.3--10~keV energy range.  Source counts were extracted from a
circular region centered on the source with a $50\arcsec$ radius, and
the background count rate was measured from a circular, source-free
area in the field of view, with a $150\arcsec$ radius.  The spectral
files were grouped to $\geq 20$ counts per energy channel.  Effective
area files corresponding to the spectral files were created using the
{\sc xrtmkarf} tool (v0.5.6), where the exposure map was taken into
account in order to correct for hot columns. Response matrices from
version 10 of the XRT calibration files were used.  The spectrum was
normalized to correspond to the 0.3--10~keV flux of the X-ray
afterglow at the epoch of the SED\@.  The normalization was determined
from the best-fit power-law decay model to the afterglow light curve,
in the same way as was done for the UVOT and ground-based data.

The SEDs were fit using {\sc XSpec} (v12.4.0), first using a single
power-law and then using a broken power-law spectral model.
In both the power-law and broken power-law models two independent dust
and gas components were included to correspond to the Galactic and the
host galaxy photoelectric absorption and dust extinction, where the
Galactic components were frozen to the column density and reddening
values taken from \citet{KBH2005} and \citet{SFD1998},
respectively. The dependence of the dust extinction on wavelength in
the GRB host galaxy was modelled on the Milky Way, the Large
Magellanic Clouds (LMC) and the Small Magellanic Clouds (SMC)
empirical extinction laws using the {\sc XSpec} model {\sc zdust},
which is based on the extinction coefficients and extinction laws from
\citet{P1992}. The total-to-selective extinction,
$R_V=A_V/E_{B\!-\!V}$ was taken to be $R_V$ = 3.08, 2.93, and 3.16 for
the Galactic, SMC and LMC extinction laws, respectively \citep{P1992}.
The equivalent neutral hydrogen column density in the host galaxy was
determined from the soft X-ray absorption, where solar abundances were
assumed.

To model the Lyman series absorption in the 912--1215~\AA\ rest-frame
wavelength range, we used the prescription provided in \citet{M1995}
to estimate the effective optical depth from the Lyman-series as a
function of wavelength and redshift, which was coded into a local
model for {\sc XSpec}. As well as estimating the hydrogen absorption
caused by intervening systems, \citet{M1995} also determined the error
on this due to statistical fluctuations in the number of absorption
clouds along the line of sight.  This error was added in quadrature to
the photometric uncertainty of any optical data at rest-frame
wavelengths blueward of Ly$\alpha$.

We found that the best-fitting models were consistent with there being
no measurable dust in the host galaxy along the line of sight to the
burst.  Since many GRB host galaxies exhibit an SMC extinction law
\citep{SFA2004,KKZ2006,SWW2007,SPO2010} we adopted this for the fits
to GRB~081029's SED\@.  However, since the amount of fitted extinction
is negligible ($A_V < 0.02$~mag at the 3-$\sigma$ level, which is
barely consistent with the upper limit derived in
\S~\ref{SECTION:xrt_data}), the details of the extinction law do not
significantly affect our results.  There is no evidence for a spectral
break between the X-ray and optical bands, so we simultaneously fit a
simple power law spectrum to all three epochs.  Our best-fitting
models for each epoch are given in Table~\ref{TABLE:sed}.  The
extinction and \ion{H}{1} column density were assumed to be the same
at every epoch.  In order to test our use of the \citet{M1995} method
of handling absorption from the intergalactic medium we removed the
ultraviolet photometry with wavelengths less than 1215~{\AA} and refit
the data.  This resulted in no significant change to the fit presented
in Table~\ref{TABLE:sed}, so we conclude that the intergalactic medium
does not significantly affect the SED of GRB~081029.

\begin{deluxetable}{cccccc}
\tabletypesize{\scriptsize}
\tablewidth{0pt}
\tablecaption{Model fits to the combined optical and X-ray SEDs.
  The $A_V$ values are 3-$\sigma$ upper limits.  The best-fit reduced
  $\chi^2$ value is $\chi^2/\nu = 92/74 = 1.24$ with a null hypothesis
  probability of 0.074.\label{TABLE:sed}}

\tablehead{%
        \colhead{Epoch} &
        \colhead{$A_V$ (mag)} &
        \colhead{$N_H$ ($10^{21}$ cm$^{-2}$)} &
        \colhead{$\beta$}}
\startdata
    4000 & $<0.02$ & 7.5  & $0.90 \pm 0.01$ \\
 12\,000 &         &      & $0.98 \pm 0.01$ \\
 20\,000 &         &      & $0.98 \pm 0.01$ \\
\enddata
\end{deluxetable}

\begin{figure}
   \centerline{
  \includegraphics[scale=0.7,angle=-90]{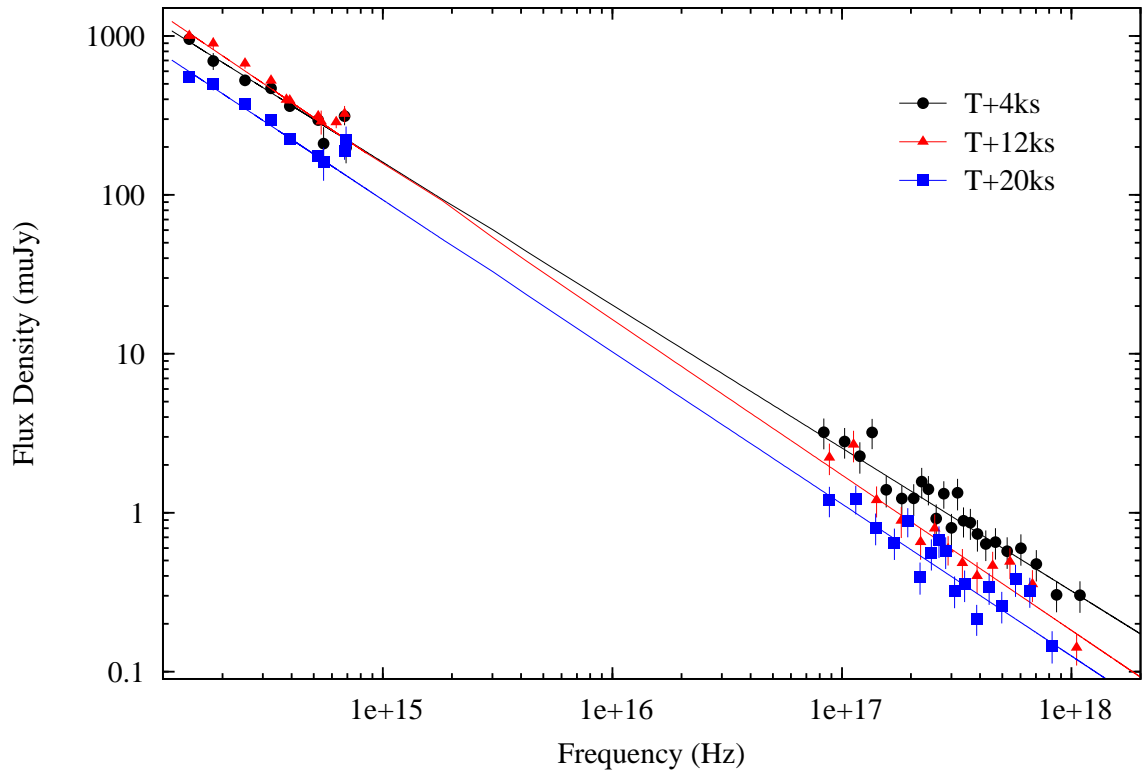}}
\caption{The Figure shows the best fit model SEDs to the optical,
  infrared, and X-ray data.  The model is the SMC broken power-law
  from Table~\ref{TABLE:sed}.  The fit at 4000~s is shown in black
  circles, the fit at 12\,000~s is shown in red triangles, and the fit
  at 20\,000~s is shown in blue squares.\label{FIGURE:sed}}
\end{figure}

The simple power-law model assumes that the optical and X-ray
photons are produced by the same mechanism.  We find that the spectrum
becomes steeper by $\Delta\beta = 0.08 \pm 0.02$ between $\sim$ 4000~s
and 12\,000~s.  This steepening occurs at about the same time that the
light curve rebrightens indicating that there is a physical change in
the mechanism that produces the light during the rebrightening.

Our SED is in agreement with the results of \citet{NGK2011}, which is
to be expected because most of our optical and near-infrared data were
taken from their paper.

\subsection{Light Curves\label{SECTION:light_curves}}

The flux density light curves for the X-ray, optical, and infrared
afterglows are presented in Figure~\ref{FIGURE:flux}.  Between about
100 and 2000~s after the BAT trigger the REM optical data decays
slowly with a decay index of $\alpha_{\mathrm{opt},1} = 0.52 \pm
0.02$.  The ROTSE $R$-band data are consistent with a smooth decay
with the same decay index.  We do not see the achromatic break at
940~s reported by \citet{NGK2011} in the GROND data.  However, the
GROND data has considerably less photometric scatter than either the
REM or ROTSE data, so this feature appears to be washed out in our
data.  There is, however, considerable variation in the $R$-band and
infrared luminosity before $\approx$500~s.  It is not clear if this
variability is real or an artefact of the photometry.  We note that
the earliest GROND infrared data show evidence for a deviation from
the smooth decay seen in the GROND optical data taken at the same
time.  This deviation is comparable to the variability seen in the
early ($t \la 500$~s) ROTSE and REM data.

Between about 3000 and 5000~s the X-ray light curve has a decay
index of $\alpha_X = 0.56 \pm 0.03$.  The optical and infrared light
curves at this time, however, rise rapidly, as is clearly seen in the
GROND data \citep{NGK2011}.  The rise index is $\alpha \approx -8$
during this period.  Our data are sparse during this period, but they
are consistent with the GROND data.  After approximately 10\,000~s our
data show the same decay as the GROND, although our data has a decay
index of $\alpha_{\mathrm{opt},2} = 1.89 \pm 0.25$ while GROND finds
$\alpha = 2.5$.  We attribute this difference the the sparsity of our
data.  The late-time X-ray decay ($\alpha_{X,2} = 2.56 \pm 0.09$) is
consistent with the late-time GROND optical decay.


\section{Interpretation\label{SECTION:interp}}

\subsection{The Synchrotron Peak\label{SECTION:synch}}

The rapid rise seen in the GROND data \citep{NGK2011} ($\alpha \approx
-8$) is inconsistent with the expected rise index of $\alpha = -0.5$
before the cooling break.  Further, the rebrightening at about 3000~s
does not show any colour evolution, so it is not possible to interpret
the rise that is seen in the optical and infrared as the synchrotron
frequency passing though the optical on its journey towards lower
energies.  Therefore, we conclude that the rise is not due to a
synchrotron peak.

\subsection{Energy Injection\label{SECTION:energy_injection}}

The break seen at 940~s in the early GROND light curves has a
magnitude of $\Delta \alpha = 0.77 \pm 0.08$ and is achromatic.  The
post-break decay index is too small for the change in decay index to
be due to a jet break, and the achromatic nature of the break argues
against it being due to the passage of the cooling break through the
optical bands.  The break can be explained by energy injection turning
off at 940~s.  If we assume a constant density interstellar medium,
and that the cooling break is above the optical bands, then having
$\alpha = 0.38 \pm 0.05$ during energy injection and $\alpha = 1.12
\pm 0.6$ after energy injection stops implies an energy injection
index of $q = 0.5 \pm 0.1$ where $L(t) \propto t^{-q}$ and an electron
distribution index of $p = 2.5 \pm 0.1$ where $N(E) \propto E^p$
\citep{DEO2009}.  Using these values we predict the spectral decay
index after the early-time break to be $\beta = 0.75 \pm 0.05$.
However, in \S~\ref{SECTION:sed} we find that the spectral decay index
between the optical and X-ray bands at 4000~s is $\beta_\mathrm{OX} =
0.90 \pm 0.01$, which is only barely consistent (a 3-$\sigma$
difference) with the expected value in the energy injection scenario.
We computed the expected spectral index for a wind-stratified
circumburst medium but were unable to find values of $p$ and $q$ that
produced a spectral decay index that was consistent with the observed
value regardless of the location of the cooling break.  The only
scenario that gives a spectral index that is roughly consistent with
the observed $\beta_\mathrm{OX} = 0.90$ is a constant density
environment with the cooling break above the X-ray band between 900~s
and 3000~s.

The energy injection scenario can explain the early-time behaviour of
the optical and infrared light curves, but it cannot explain the
rebrightening seen at 3000~s.  We need some other mechanism to do
this.

\subsection{A Two-Component Jet\label{SECTION:2jet}}

We are able to reproduce most of the observed X-ray, optical, and
infrared light curves, as well as the observed SED if we assume a
two-component jet model for the afterglow of GRB~081029.  The
afterglow is characterized by a rebrightening in the optical and
infrared bands with a simultaneous flattening in the X-ray band.  This
implies that a new mechanism was contributing to the flux in the
optical regime starting at about 3000~s.  In the two-component jet
model, the early afterglow emission ($t < 2500$~s) was produced by the
narrow, fast component while the late rebrightening was attributed to
the emergence of the radiation powered by the wider, slower component.

In our fit the deceleration time for the wide component occurs earlier
than 3000s, and we find that the synchrotron frequency of the wide jet
component of the afterglow is between the X-ray and optical bands
(\ie, $\nu_\mathrm{opt} < \nu_\mathrm{m,w} < \nu_\mathrm{X}$) for
$3000 < t < 9000$~s.  The passage of the wide jet's synchrotron break
through the optical band cannot reproduce the rapid rise that is seen
in the optical and infrared photometry at about 3000 s, suggesting
that there is another process at work that contributes to the sudden
increase in the flux.  Further,~\cite{N2011} find that optical flux
rises as $t^{-8}$ during this time.

The physical parameters of the two components are summarized in
Table~\ref{TABLE:model}.  The half-opening angle of the jet is denoted
by $\theta_j$, $\Gamma_0$ is the Lorentz factor, $E_{K,\mathrm{iso}}$
is the isotropic equivalent kinetic energy in the jet, $p$ is the
electron index, $\epsilon_e$ and $\epsilon_B$ are the fractions of the
energy in electrons and magnetic fields respectively, $n$ is the
density of the circumburst medium, and $z$ is the redshift.  Details
of the model and the numerical code used are given in \citet{JYF2007}.

\begin{figure}
  \plotone{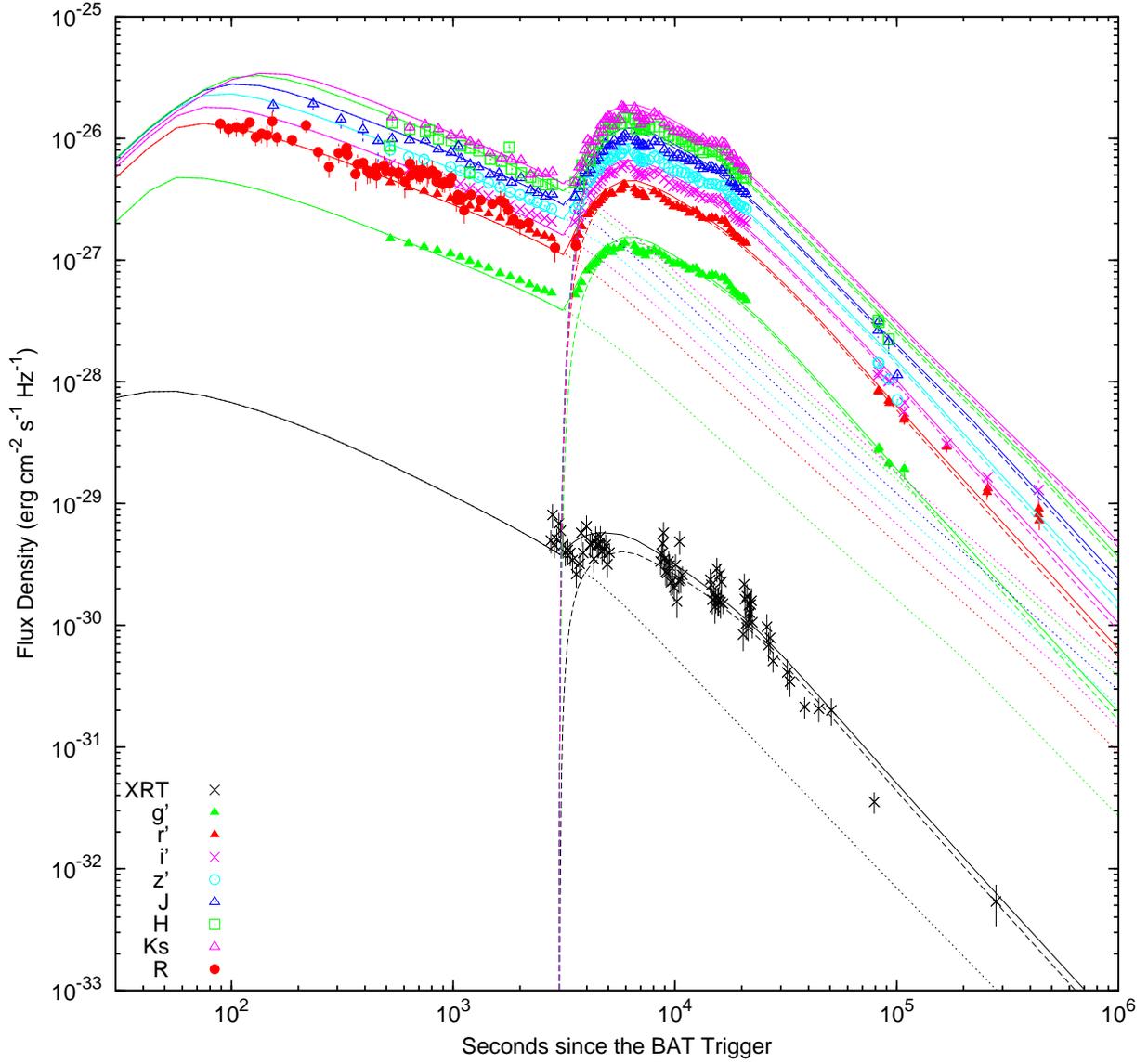} \figcaption[./zhiping_model.eps]{This
    Figure shows the best-fitting two-component jet model for our
    light curves.  The colours used in the Figure are the same as in
    Figure~\ref{FIGURE:flux}.\label{FIGURE:model}}
\end{figure}

\begin{deluxetable}{ccc}
\tabletypesize{\scriptsize}
\tablewidth{0pt}
\tablecaption{Model fits for a two-component jet.\label{TABLE:model}}
\tablehead{%
        \colhead{Parameter} &
        \colhead{Narrow Jet} &
        \colhead{Wide Jet}}
\startdata
$\theta_j$ (rad)         & 0.015               & 0.025 \\
$\Gamma_0$               & 500                 & 100 \\
$E_{K,\mathrm{iso}}$ (erg) & $4.0 \times 10^{54}$ & $3.0 \times 10^{54}$ \\
$p$                      & $2.05$              & $2.20$ \\
$\epsilon_e$             & 0.05                & 0.10 \\
$\epsilon_B$             & 0.0001              & 0.0002 \\
$n$ (cm$^{-3})$          & 10                  & 10 \\
$z$                      & 3.8479              & 3.8479 \\
\enddata
\end{deluxetable}

We find that the narrow, inner jet has a half-opening angle of
$\theta_{j,n} = 0.015$~rad ($0\fdg9$) and an initial Lorentz factor of
$\Gamma_{0,n} = 500$.  This component gives rise to the X-ray flux
and the pre-jump optical and infrared flux.  The wider, outer jet has
$\theta_{j,w} = 0.025$~rad ($1\fdg4$) and an initial Lorentz factor of
$\Gamma_{0,w} = 100$.  This component dominates the afterglow after
about 3000~s.  The total electromagnetic energy in the afterglow is
approximately equally divided between the two jets.

Our two-component model predicts that the optical spectrum of the wide
jet emission (which dominates at late time) has $\beta_\mathrm{opt} =
0.6$ and the X-ray spectrum has $\beta_X = 1.1$, with the break
being located at $\sim 10^{15}$ Hz (approximately the $u$ band), at
$\sim 9000$~s
This is not consistent with the fits to the observed SEDs presented in
\S~\ref{SECTION:sed}.  We tried varying the amount of extinction when
fitting our models and found that the observed SEDs can be made to
weakly agree with the model if there is extinction in the host galaxy
along the line of sight to the burst.  This is consistent with the
constraints on the extinction ($A_V \lesssim 2$~mag in the host
galaxy) from the X-ray data alone (\S~\ref{SECTION:xrt_data}) but
inconsistent with the stronger constraint on the host extinction ($A_V
< 0.03$~mag) found by combining the X-ray, optical, and infrared
data (\S~\ref{SECTION:sed}).

As shown in Figure~\ref{FIGURE:model} the two-component jet model can
reproduce the X-ray and some of the optical/infrared data reasonably
well.  However, this model fails to reproduce the rapid rise seen in
the UVOT white data.  Further, GROND observations show that this rapid
rise occurs in all filters between the GROND $g\prime$ and $K_s$ bands
and has a power law index of $\alpha \sim -8$ \citep{N2011}.  This is
somewhat steeper than can be accommodated with the two-component jet
model.  Finally, the two-component model predicts either a spectral
break at about $10^{15}$~Hz or extinction along the line of sight in
the host galaxy.  Both of these predictions are inconsistent with the
data.  The two-component jet model can explain the observed light
curves of the afterglow, but not the spectral behaviour.  Further, the
two-component jet model has trouble handling the rapid rise in the
flux seen at $\approx$3000~s.

\subsection{Colliding Shells\label{SECTION:shells}}

\citet{VVM2011} have proposed that collisions between ejecta shells
can produce flares in the optical light curve.  In their scenario two
shells are ejected by the central engine.  The first shell has a lower
Lorentz factor than the second shell, so the second shell will
eventually catch up with the first.  The first shell sweeps up a
uniform interstellar medium and decelerates.  The second shell has a
higher Lorentz factor and overtakes the first shell.  The collision
between the two shells produces an optical flare with properties that
depend on the Lorentz factor of the second shell and the isotropic
energy ($E_\mathrm{iso}$).  \citet{VVM2011} find that the collision
between two shells of material with different Lorentz factors can
produce an optical flare with $\Delta t / t \sim 1$.  This is
approximately consistent with the GROND data for the rebrightening
episode.  The simulations of \citet{VVM2011} suggest that the
magnitude of the flare, relative to the underlying synchrotron light
curve, depends on the Lorentz factor and the isotropic energy.  They
find typical values for the increase in the flux ($f$) resulting from
the collision of $\Delta f / f \sim 2$--5 for typical GRB values of
$\Gamma$ and $E_\mathrm{iso}$.  This is consistent with what is seen
during the rebrightening of GRB~081029.

Figures~4~and~5 of \citet{VVM2011} show predicted optical light curves
produced by colliding ejecta shells for four sets of Lorentz factors
and isotropic energies.  All four cases result in light curves that
exhibit flares that have shapes and intensities that are similar to
the rebrightening seen in GRB~081029.  The simulations assume $\Gamma
= 23$ for the first shell, which results in the onset of a flare at
$\sim 20\,000$~s in the rest frame.  Our data suggest that the
GRB~081029's flare started at $\sim 3000$~s in the observer's frame
($\sim 600$~s in GRB~081029's rest frame).  However, the time of the
collision will depend on the time that the second (faster) shell was
ejected relative to the first (slower) shell, and on the distance of
the first shell from the central engine (and the first shell's Lorentz
factor).  Detailed simulations will be needed to test this scenario
and determine the physical properties of the ejecta.


\section{Discussion\label{SECTION:discussion}}

\subsection{The Afterglow}\label{SECTION:afterglow}

In general we find that by themselves neither a one-component jet nor
continuous energy injection from the central engine can explain the
observed light curves and SED of the X-ray, optical, and infrared
afterglows of GRB~081029.  A two-component jet model, similar to what
is seen in some other GRB afterglows, does provide a reasonable fit to
the light curves, but the Lorentz factor of the fast, narrow jet is
less than expected given that the peak time of the light curve is
earlier than 89~s in the observer's frame.  Further, the two-component
jet model is not able to reproduce the observed SEDs during the
unusual optical activity.

We find that the rise in the optical light curves of the afterglow of
GRB~081029 can be broadly explained by the collision of a fast-moving
ejecta shell with a slower shell that has been decelerated by sweeping
up a uniform interstellar medium.  This scenario does not, however,
address the shallow decay phase of the afterglow.  The early shallow
decay requires that the emission is due to a different emission
component from the late-time emission.  We favour the multi-component
jet explanation because it does not require energy injection.  The
discrepancy between the observed SED (with $\beta = 1$ and $A_V <
0.05$~mag) and the SED that is predicted by by the two-jet model (with
$\beta = 0.6$ and $A_V \sim 0$~mag) can be explained by the spectrum
at $\sim~12\,000$~s being dominated by emission from the flare caused
by the collision between the two shells.  Evidence for this is that at
4000~s, during the onset of the flare, the SED had $\beta \approx
1.0$, which suggests a transition between the intrinsic SED of the
narrow jet and the SED of the emission from the collision.  If this is
the case there is no need to invoke extinction to explain the
difference between the observed and predicted spectral decay indices.
This leads to a picture where GRB~081029 had a two-component jet and a
collision between two ejecta shells at about 3000~s.  At this time the
afterglow is making a transition from being dominated by the narrow
jet to being dominated by the wide jet, so it is not possible to tell
if the collision between the two ejecta shells occurred in the narrow
or wide jet.

The nature of GRB afterglows has been a matter of much debate over the
past decade.  There is a general agreement that they are the result of
a combination of a forward shock due to a relativistic jet moving into
the circumstellar medium surrounding the burst and a reverse shock
that propagates back into the jet.  However, the details of how these
shocks affect their environment, the role of magnetic fields, and the
structure of the jets are the subject of much research.  Several GRBs
have had afterglows that are difficult or impossible to explain using
a single, uniform jet.  A multi-component jet structure has been
postulated to explain unusual behaviour in the light curves of some
GRBs.  An example of a multi-component jet is given by
\citet{BKP2003}, who invoked a two-component jet to explain radio
observations of the long--soft burst GRB~030329.  \citet{ODP2007}
found that a two-component jet explained GRB~050802's afterglow, and
\citet{HBG2007} found that a two-component jet could explain the lack
of a jet break in the light curves of XRF~050416A.  \citet{RKS2008}
found that the afterglow emission from the ``naked-eye burst''
GRB~080319B is best explained using a two-component jet.  A
multi-component jet can also explain the afterglow of the short--hard
burst GRB~051221A \citep{JYF2007}.  However, the physical parameters
of multi-component jets vary considerably from one GRB to another, so
there does not appear to be a universal jet structure.

Multi-component or structured jets are predicted by simulations of the
relativistic outflow from GRBs.  \citet{KG2003} found that the bulk
Lorenz factor decreases as one moves away from the axis of the jet
resulting in a jet with a fast inner core surrounded by a slower outer
envelope.  Simulations of outflows from accretion discs about
collapsed massive stars show that multi-component jets can form with
the outer jet carrying far more energy than the inner jet
\citep{VPK2003}.  In this scenario the inner and outer jet will have
different energies and different bulk Lorentz factors.  The
interaction of each jet with the circumburst medium will produce
separate afterglow emission components, which can result in complex
light curves \citep{PKG2005}.

Several mechanisms have been proposed to create late-time activity in
the central engine of a GRB that would result in multiple ejecta
events.  \citet{KOG2005} suggest that the fragmentation of a
rapidly-rotating stellar core could result in multiple accretion
events onto the newly-formed compact object.  \citet{PAZ2006} pointed
out that the fragmentation of an accretion disc that is undergoing
viscous evolution can result in an accretion disc that results in
highly variable accretion onto the central compact object.  Each
accretion event would restart the central engine resulting in new
shells being ejected.  Each accretion event would be independent, so
the initial Lorentz factors of the ejecta could vary considerably
leading to collisions between shells from different events.

There is evidence for colliding shells in the optical light curves of
a few GRBs.  The afterglow of GRB~081029 has an optical light curve
that is similar to that of GRB~060206 \citep{SDP2007,WVW2006}, and GRB
970508 \citep{SKZ1998} exhibited a late-time flare similar to what is
expected from colliding shells.  After ten years such flares have only
been observed in a handful of GRB afterglows, which suggests that
discrete late-time accretion events may be fairly uncommon in GRBs.

A combination of the colliding shell scenario and a multi-component
jet can reproduce the broad features of the light curves, but detailed
modelling will be needed to determine the physical parameters
governing this afterglow.  Several GRB afterglows have shown evidence
for multi-component jets, and there is evidence that some GRBs undergo
multiple accretion events that result in late-time impulsive energy
injection into the afterglow.  GRB afterglows appear to be complex
phenomena that require detailed modelling to be fully understood.

\subsection{The Host Galaxy}\label{SECTION:host}

We detect both hydrogen absorbing features (Ly$\alpha$ and Ly$\beta$)
and several metallic transitions in the spectrum of the optical
afterglow.  The latter belong both to neutral elements
(\ion{O}{1}($\lambda$1039) and ($\lambda$1302)), low ionization
species (\ion{C}{2}($\lambda$1334), \ion{S}{2}($\lambda$1250),
($\lambda$1253), and ($\lambda$1259), \ion{Si}{2}($\lambda$1264),
($\lambda$1304), and ($\lambda$1526), \ion{Fe}{2}($\lambda$1608),
\ion{Al}{2}($\lambda$1670), \ion{Al}{3}(1854)) and high ionization
species (\ion{N}{5}($\lambda$1238) and ($\lambda$1242),
\ion{Si}{4}($\lambda$1393) and ($\lambda$1402),
\ion{C}{4}($\lambda$1548) and ($\lambda$1550),
\ion{O}{6}($\lambda$1031) and ($\lambda$1037)).  In addition to these
features, lines from several fine structure levels of
\ion{C}{2}($\lambda$1334), \ion{O}{1}($\lambda$1304) and
($\lambda$1306), \ion{Si}{2}($\lambda$1264), ($\lambda$1309), and
($\lambda$1533), and \ion{Fe}{2}($\lambda$1618), ($\lambda$1621),
($\lambda$1631), and ($\lambda$1636) are detected.  These features are
excited by the ultraviolet flux from the GRB afterglow.  Estimates of
the typical distance from a GRB to absorption systems suggest
distances of $\sim 0.1$--1~kpc \citep{VLS2007,DFP2009} implying that
all these absorption features are due to the GRB~081029 host galaxy.
The common redshift of these features is $z=3.848$, which we take at
the redshift of the host.  We find no evidence for intervening metal
absorption lines in our combined spectrum.


\section{Conclusions\label{SECTION:conc}}

GRB~081029 was a long--soft GRB with a redshift of $z = 3.8479$.  It
had a smooth gamma-ray light curve and did not appear to have any
unusual gamma-ray properties.  Neither the gamma-ray nor the X-ray
properties of this burst showed any sign of strange behaviour.  The
optical and infrared light curves, on the other hand, were not typical
of GRB afterglows.  There was a brightening in the optical and
infrared light curves at about 3000~s that cannot be explained as the
passage of the synchrotron break through the optical, by a
two-component jet model, or by continuous energy injection from the
central engine.  We find that the combination of the colliding shell
scenario of \citet{VVM2011} and a two-component jet can reproduce the
unusual optical light curve of this afterglow.  Our result is
consistent with a central engine that was reactivated by a discrete,
major accretion event.


\acknowledgements

We acknowledge the use of public data from HEASARC's {\sl Swift\/}
Data Archive.  The ROTSE project is supported by the NASA grant
NNX08AV63G and the NSF grant PHY-0801007.  The authors wish to thank
Scott Barthelmy and the GRB Coordinates Network for rapidly providing
precise GRB positions to the astronomical community.  This research
has made use of the NASA/IPAC Extragalactic Database, which is
operated by the Jet Propulsion Laboratory, California Institute of
Technology, under contract with NASA.  The authors would like to thank
the anonymous referee for a thorough review of this paper.


  
\end{document}